\documentclass[12pt]{iopart}
\usepackage{citesort}
\usepackage{graphicx}
\begin{document}
\title{Spin fluctuations in the 112-type iron-based superconductor Ca$_{0.82}$La$_{0.18}$Fe$_{0.96}$Ni$_{0.04}$As$_{2}$}

\author{Tao Xie$^{1,2,3*}$, Chang Liu$^{1,2*}$, Ryoichi Kajimoto$^4$, Kazuhiko Ikeuchi$^5$,  Shiliang Li$^{1,2,6}$ and Huiqian Luo$^{1,6\dag}$}

\address{$^1$ Beijing National Laboratory for Condensed Matter Physics, Institute of Physics, Chinese Academy of Sciences, Beijing 100190, China}
\address{$^2$ School of Physical Sciences, University of Chinese Academy of Sciences, Beijing 100190, China}
\address{$^3$ Neutron Scattering Division, Oak Ridge National Laboratory, Oak Ridge, Tennessee 37831, USA}
\address{$^4$ Materials and Life Science Division, J-PARC Center, Japan Atomic Energy Agency, Tokai, Ibaraki 319-1195, Japan}
\address{$^5$ Neutron Science and Technology Center, Comprehensive Research Organization for Science and Society,
Tokai, Ibaraki 319-1106, Japan}
\address{$^6$ Songshan Lake Materials Laboratory, Dongguan, Guangdong 523808, China}

\ead{$^{*}$These authors contributed equally to this work.}
\ead{$^{\dag}$hqluo@iphy.ac.cn}

\begin{abstract}
We report time-of-flight inelastic neutron scattering (INS) investigations on the spin fluctuation spectrum in the 112-type iron-based superconductor (FeSC) Ca$_{0.82}$La$_{0.18}$Fe$_{0.96}$Ni$_{0.04}$As$_{2}$ (CaLa-112). In comparison to the 122-type FeSCs with a centrosymmetric tetragonal lattice structure (space group $I4/mmm$) at room temperature and an in-plane stripe-type antiferromagnetic (AF) order at low temperature, the 112 system has a noncentrosymmetric structure (space group $P2_{1}$) with additional zigzag arsenic chains between Ca/La layers  and a magnetic ground state with similar wavevector $\mathbf{Q}_{\mathrm{AF}}$ but different orientations of ordered moments in the parent compounds. Our INS study clearly reveals that the in-plane dispersions and the bandwidth of spin excitations in the superconducting CaLa-112 closely resemble to those in 122 systems. While the total fluctuating moments $\langle m^2 \rangle\approx 4.6\pm0.2 \mu_B^2$/Fe are larger than 122 system, the dynamic correlation lengths are similar ($\xi\approx 10$ \AA). These results suggest that superconductivity in iron arsenides may have a common magnetic origin under similar magnetic exchange couplings with a dual nature from local moments and itinerant electrons, despite their different magnetic patterns and lattice symmetries.

\end{abstract}


\maketitle
\ioptwocol

\section{Introduction}

Unconventional superconductivity usually emerges from the antiferromagentic (AF) parent compounds such as cuprates, nickelates, iron pnictides and chalcogenides ~\cite{xgwen2006,xjzhou2021,qgu2022,xhchen2014,qsi2016,bdwhite2016,wwu2014,jgcheng2015,ptyang2022,kamihara2008}.  By introducing charge carrier doping or applying external pressure to induce superconductivity, the long-range static AF orders in the parent compounds are gradually suppressed, but short-range dynamic AF fluctuations persist throughout the superconducting region and are coupled directly with the occurrence of superconductivity ~\cite{stewart2011,pdai2012,jttranquada2014,pdai2015,jhnson2015,gong2018,hqluo2017,tzhou2020,hqluo2012,luxy2013,dhu2015,qgao2020}. Thus AF fluctuations are commonly considered as the pairing glue of the Cooper pairs in unconventional superconductors ~\cite{jttranquada2014,pdai2015,gong2018,hqluo2017,jhnson2015}. Within this picture, the change of the magnetic exchange energy ($\Delta E_{ex}$) below and above the superconducting transition temperature $T_c$ should account for the superconducting condensation energy ($U_c$) ~\cite{ostockert2011,dsinosov2010,mwang2013,dhu2016}. Theoretically, it is defined by: $\Delta E_{ex}=\Delta \sum_{ij} J_{ij}\langle \mathbf{S}_{i}\cdot\mathbf{S}_{j}\rangle$, where $J_{ij}$ is the magnetic exchange coupling and $\langle \mathbf{S}_{i}\cdot\mathbf{S}_{j}\rangle$ is the dynamical spin susceptibility in absolute units ~\cite{mwang2013,scalapino2012}. Therefore, it is essential to experimentally determine both the local magnetic interactions and the absolute strength of magnetic fluctuations in unconventional superconductors, which can be measured by inelastic neutron scattering (INS) ~\cite{jttranquada2014,pdai2015,jhnson2015,gong2018}.

While it is a great challenge for INS to map the complete spectrum of spin fluctuations in high-$T_c$ cuprates due to their large energy scale of bandwidth over than 300 meV ~\cite{jttranquada2014,mfujita2012,nsheadings2010,smhaydena1996,rcoldea2001}, the iron-based superconductors (FeSCs) provide excellent opportunities to extensively compare the spin fluctuations among different systems due to the relatively lower bandwidth around 200 meV and available large crystals for different families with various kinds of doping~\cite{pdai2015,jhnson2015,gong2018,hqluo2017,jzhao2009,raewings2011,lwharriger2011,msliu2012,hqluo2013,xylu2018,hfli2010,gstucker2012,dhu2016,mdlumsden2011,ojlipscombe2011,mywang2011}. Among the FeSCs studied by INS, the 122 systems including (Ba, Sr, Ca)Fe$_2$As$_2$ and the related doped compounds, have a well established picture for the doping dependence of the spin fluctuations~\cite{jzhao2009,raewings2011,lwharriger2011,msliu2012,hqluo2013,xylu2018,hfli2010,gstucker2012}. Taking BaFe$_2$As$_2$ as an example, while the electron dopings Ni to Fe sites only suppress the spin excitations below 100 meV ~\cite{lwharriger2011,msliu2012,hqluo2013,xylu2018,hfli2010,gstucker2012}, the hole dopings K to Ba sites strongly suppress the high energy part instead~\cite{mwang2013}. On the other hand, the isovalent dopings P to As sites increase the effective bandwidth of spin excitations ~\cite{dhu2016}. In spite of the doping dependent strength and the dramatically different dispersions of spin fluctuations for various iron-based materials, the next-nearest-neighbor (NNN) effective coupling in their parent compounds are generally AF and rather similar (about 20 meV). Additionally, the total fluctuating moments $\langle m^2 \rangle\approx 3 \mu_B^2$ per Fe are similar to cuprates~\cite{mwang2013,pdai2015,jhnson2015,msliu2012}.  More importantly, a neutron spin resonance mode served as the hallmark of magnetically driven electron pairing is extensively observed in many FeSCs ~\cite{adchristianson2008,yqiu2009,nqureshi2012,czhang2013,mwang2010,czhang2011,chlee2013,jtpark2011,qwang2016,mma2017,txie2018a,txie2018b,wshong2020,txie2021,cliu2022}, and the spin fluctuations in the superconducting compounds are strong enough to make $\Delta E_{ex}>>U_c$ ~\cite{mwang2013,dhu2016,pdai2015,jhnson2015}.

\begin{figure}[tbp]
\center
\includegraphics[width=0.5\textwidth]{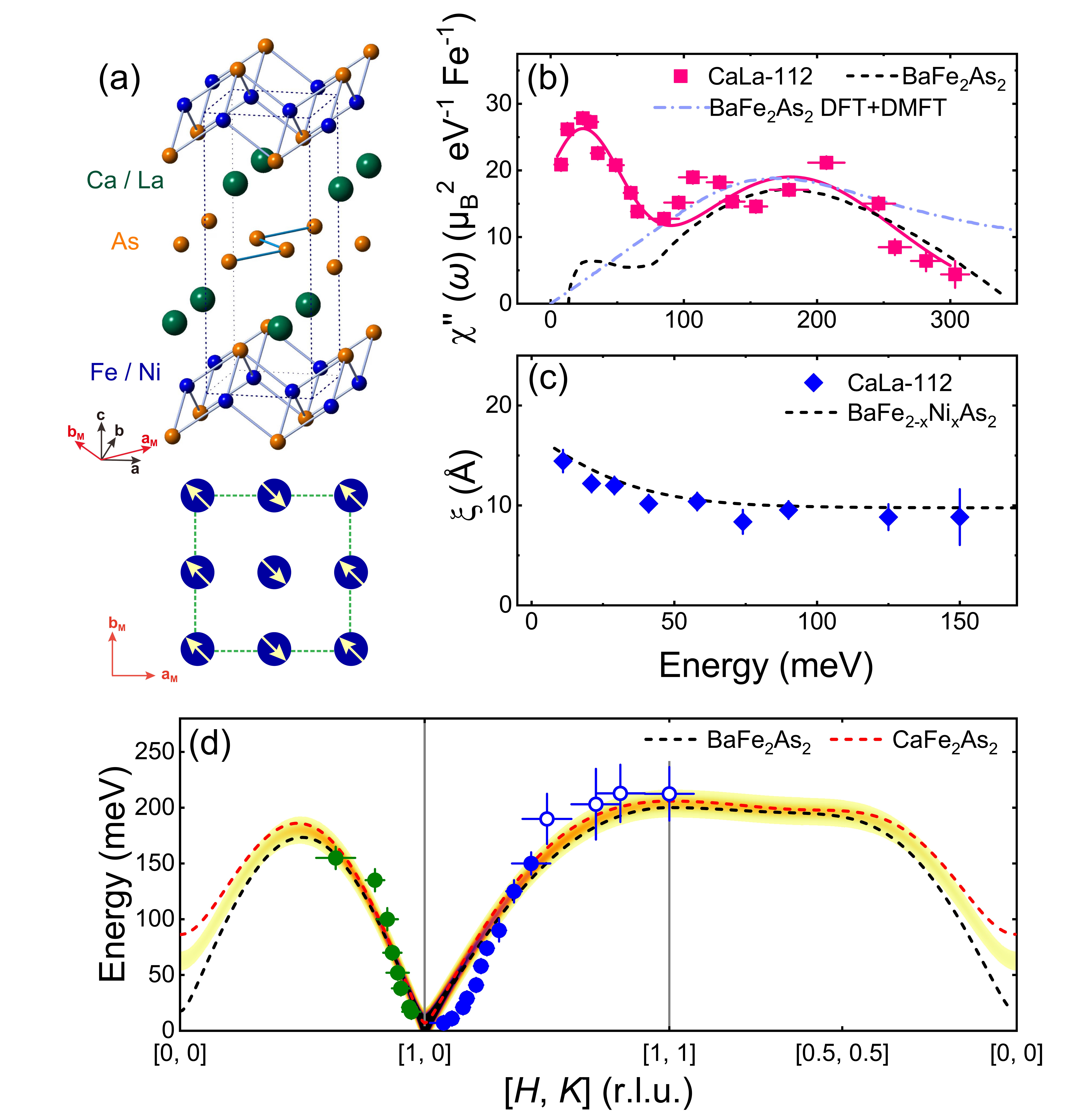}
\caption{(a) Crystal structure (above) and magnetic structure (below) of Ca$_{1-y}$La$_{y}$FeAs$_{2}$ system. The magnetic unit cell marked by $\mathbf{a}_\mathrm{M}$ and $\mathbf{b}_\mathrm{M}$ axes is used to define the reciprocal space in our experiment. (b) Energy dependence of the local dynamic susceptibility $\chi^{\prime\prime}(E)$ for CaLa-112 in the units of $\mu_B^2$/eV/Fe. The solid line is a guide to eyes, the dashed line is the $\chi^{\prime\prime}(E)$ for BaFe$_{2}$As$_{2}$, and the dashed dot line is the DFT+DMFT result for BaFe$_{2}$As$_{2}$  ~\cite{lwharriger2011,msliu2012,hqluo2013}. (c) Energy dependence of the dynamic spin-spin correlation lengths $\xi$ for CaLa-112. (d) The in-plane dispersions of spin excitations along high symmetric directions: [0, 0]-[1, 0]-[1, 1]-[0, 0]. The solid dots are obtained from two-gaussian peak fitting of the $H$ or $K$-cuts at constant-energy window, and the open circles are obtained from the peak fitting of $E$-cuts at fixed $K$ points. The dashed lines are spin wave dispersions in the 122-type parent compound CaFe$_{2}$As$_{2}$ and BaFe$_{2}$As$_{2}$, and the gradient colors are guides to eyes.}
\label{fig1}
\end{figure}

Here, we report INS investigations on the spin fluctuation spectrum in the 112-type FeSC Ca$_{0.82}$La$_{0.18}$Fe$_{0.96}$Ni$_{0.04}$As$_{2}$ (CaLa-112) with $T_c=22$ K. Different from those 122/1111/111 systems with a tetragonal lattice structure (space group I4/mmm or P4/mmm ) at room temperature and an in-plane stripe-type AF order at low temperature, the 112 system has a unique noncentrosymmetric structure (space group $P2_{1}$) with additional zigzag arsenic chains between Ca/La layers and a magnetic ground state with a similar wavevector $\mathbf{Q}_{\mathrm{AF}}$ but different orientation of magnetic moments in the parent compounds, which is 45$^{\circ}$ away from the stripe direction [Fig. 1(a)] ~\cite{nkatayama2013,sjiang2016a,sjiang2016b,txie2017,jyu2021,cyjiang2020}. Our previous study on the low energy spin excitations in the same compound revealed a two-dimensional spin resonance mode under weak spin-orbit coupling ~\cite{txie2018a}. Using time-of-flight INS technique, we are able to further map out the high-energy spin excitations in CaLa-112 and compare with the 122 systems. As shown in Fig. 1 (b)-(d), the fluctuating strength $\chi^{\prime\prime}(E)$ is initially enhanced at low energy then decreases to a minimum around 70 meV before increasing again to a maximum around 180 meV. The in-plane spin-spin correlation lengths at all energies are similar to that in BaFe$_{2-x}$Ni$_x$As$_2$ ($\xi\approx 10$ \AA) ~\cite{lwharriger2011,msliu2012,hqluo2013}. Moreover, the overall dispersions within the FeAs-plane and the energy bandwidth are also similar to BaFe$_2$As$_2$ and CaFe$_2$As$_2$ ~\cite{jzhao2009,lwharriger2011}, suggesting that the magnetic exchange couplings are similar in these two different systems.

\begin{figure*}[tbp]
\center
\includegraphics[width=1\textwidth]{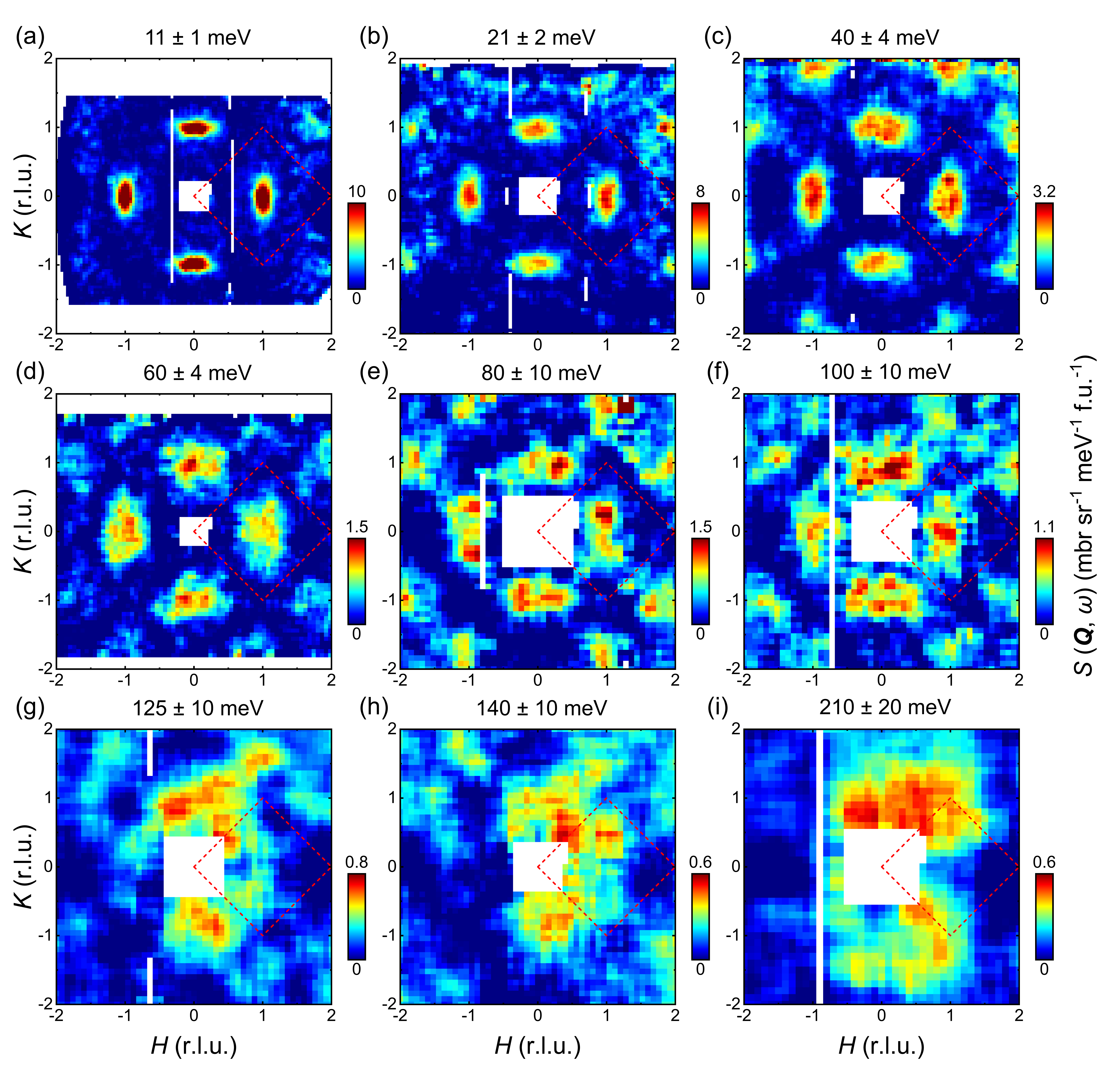}
\caption{Two dimensional constant-energy slices through the magnetic excitations of CaLa-112 at energies of $E=11\pm 1$, $21\pm 2$, $40\pm 4$, $60\pm 4$, $80\pm 10$, $100\pm 10$, $125\pm 10$, $140\pm 10$ and $210\pm 20$ meV.  The data in (a), (b), (c,d), (e,f,g,h) and (i) are collected using $E_i= 42, 73, 99, 250, 449$ meV, respectively. The data sets below 75 meV are subtracted by a radially symmetric $Q$-dependent background integrated from the diagonal line of the entire zone $-2<H<2$ and $-2<K<2$, which is mainly from the phonon scattering of the aluminum sample holders. The data sets above 75 meV are subtracted by a background integrated from $1.8<H<2.2$ and $-0.2<K<0.2$, which is from the incoherent scattering.  The color bars represent the vanadium normalized absolute spin excitation intensity in the units of mbarn/sr/meV/f.u. and the dashed red boxes indicate the integration region for calculating energy-dependent local dynamic susceptibility.
}
\label{fig2}
\end{figure*}

\section{Experimental details}
We used the same sample set of CaLa-112 crystals in our previous INS experiments ~\cite{txie2018a}. These crystals were grown by self-flux method, and its composition (Ca$_{0.82}$La$_{0.18}$Fe$_{0.96}$Ni$_{0.04}$As$_{2}$) were determined by the inductively coupled plasma analysis ~\cite{txie2017}. About 2.3 grams of single crystals ($\sim$1500 pieces) were co-aligned in the scattering plane $[H, 0, 0] \times [0, 0, L]$ defined by ${\bf Q}=(H,K,L) = (q_xa/2\pi, q_yb/2\pi, q_zc/2\pi)$ in reciprocal lattice unit (r.l.u.) using the pseudo-orthorhombic magnetic unit cell with $a_M\approx b_M\approx 5.54$ \AA, $c_M=10.27$ \AA\ [Fig. 1(a)]. The mosaics spread of the co-aligned assembly were about 3.7 degrees in $ab$-plane and 2.8 degrees for out-of-plane case ~\cite{txie2018a}. Time-of-flight INS experiments were carried out at 4SEASONS chopper spectrometer (BL-01) at J-PARC, Tokai, Japan, with multiple incident energies $E_i$ = 449, 250, 99, 73, 42 meV, $k_i$ parallel to the $c$-axis, and chopper frequency $f=$ 250 Hz ~\cite{nakamura2009,kajimoto2011}. All data were collected at base temperature $T=5$ K and analyzed by the Utsusemi, DAVE and Horace software packages \cite{inamura2013,rtazuah2009,raewings2009}. To compare easily with spin waves in BaFe$_{2}$As$_{2}$ and CaFe$_{2}$As$_{2}$, all data were normalized to the absolute units (mbarn/sr/meV/f.u.) using a vanadium standard method ~\cite{jzhao2009,lwharriger2011,msliu2012,hqluo2013,gyxu2013}. The INS directly measured the differential scattering cross section ~\cite{mwang2013}:
$\frac{d^2\sigma}{d\Omega dE}\frac{k_i}{k_f}=\frac{2(\gamma r_e)^2}{\pi g^2\mu_{B}^{2}}|F(\mathbf{Q})|^{2} \frac{\chi^{\prime\prime}(Q,E)}{1-\exp(-E/k_BT)}.$ Here $F(\mathbf{Q})$ is the magnetic form factor of Fe$^{2+}$, $1/[1-\exp(-E/k_BT)]$ is the Bose population factor, $k_B$ is the Boltzmann constant, and $\chi^{\prime\prime}(Q,E)$ is the imaginary part of the dynamic susceptibility. After subtracting the background and correcting the form factor and Bose factor, the data were multiplied by the constant $\pi g^2\mu_{B}^{2}/(2(\gamma r_e)^2)=21.6289$ to covert into the absolute units of $\mu_B^2$/eV/f.u..  We calculated the local dynamic susceptibility using $\chi^{\prime\prime}(E)=\int{\chi^{\prime\prime}({\bf Q},E)d{\bf Q}}/\int{d{\bf Q}}$ , where $\chi^{\prime\prime}({\bf Q},E)=(1/3) tr( \chi_{\alpha \beta}^{\prime\prime}({\bf Q},E))$. The total fluctuating moments can be obtained by further integrating on energy:\\
$\langle m^2 \rangle=(3/\pi)\int{\chi^{\prime\prime}(E)/[1-\exp(-E/k_BT)]dE}$ \\ ~\cite{lwharriger2011,msliu2012,hqluo2013}. It should be noticed that there is only one Fe atom in the unit cell for the single-layer CaLa-112 system but two Fe atoms in double-layer 122-compound, so we compare $\chi^{\prime\prime}(E)$ in the units of $\mu_B^2$/eV/Fe for clarity [Fig. 1(b)]. In principle, the absolute magnitude of magnetic scattering can be also estimated by normalizing it to phonon intensities ~\cite{gyxu2013}. Unfortunately, there were very limited phonon signals could be found in our measurements. Due to the setup of $k_i \parallel c$, the energy transfer $E$ and momentum transfer along $L$ direction were in coupled with each other, which made it was difficult to observe the full phonon signals in the three dimensional reciprocal space. Considering the large uncertainty in such estimation by phonon intensities, we preferred to use the vanadium standard method for normalization as we extensively did before in other iron-based superconductors ~\cite{lwharriger2011,msliu2012,hqluo2013}.

\begin{figure*}[tbp]
\center
\includegraphics[width=0.9\textwidth]{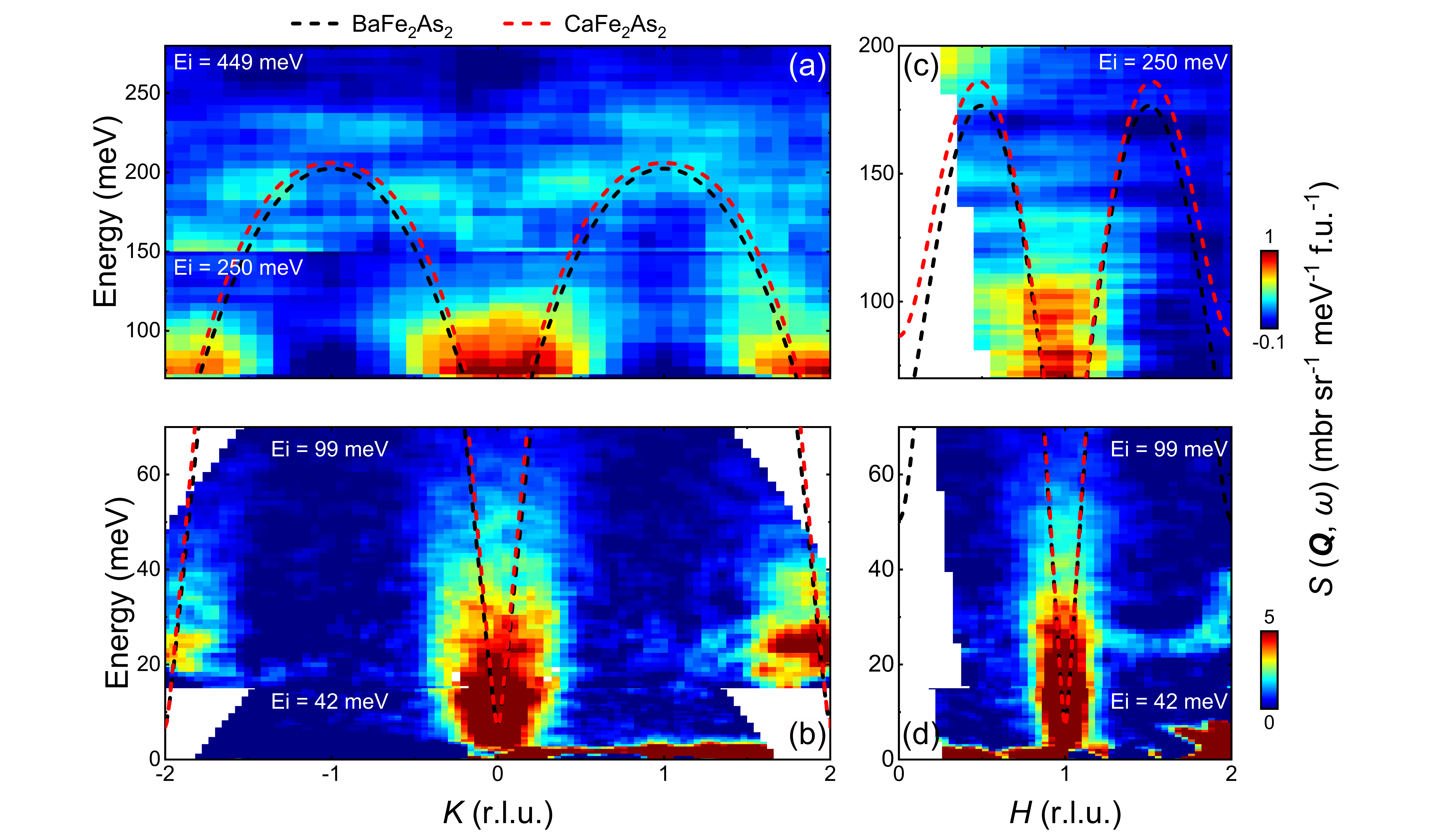}
\caption{Energy dependence of the two-dimensional slices along the $[1, K]$ and $[H, 0]$ directions with $E_i= 450$, 250, 99 and 42 meV for panels (a,c) and (b,d), respectively.
The black and red dashed lines are dispersions of spin waves in CaFe$_{2}$As$_{2}$ and BaFe$_{2}$As$_{2}$.}
\label{fig3}
\end{figure*}

\section{Results and discussions}

In principle, the obtained data sets are four-dimensional (4D) in the three-axes reciprocal space $[H, K, L]$ plus an energy axis $E$. Since the constant-energy spin excitations are nearly two-dimensional (2D) in reciprocal space ~\cite{txie2018a}, and the energy transfer $E$ actually couples with $L$ due to $k_i \parallel c$ setup, we present our data details mainly in $[H, K]$ plane for different energy windows, as shown in Figs. 2 - 4. The 2D constant-energy slices of spin excitations of CaLa-112 in the $[H, K]$ plane are shown in Fig. 2. The scattering intensity is normalized to absolute units of mbarn/sr/meV/f.u. using a vanadium standard. The dashed boxes mark the integration region for calculating energy-dependent local dynamic susceptibility, which is equivalent to the AF Brillouin zone for the magnetic unit cell with single Fe$^{2+}$. As we measured the electron overdoped system, there is no spin gap due to $J_c=J_s=0$, which is commonly observed in the 3D spin waves of BaFe$_{2}$As$_{2}$ and CaFe$_{2}$As$_{2}$ ($E_g=10 \sim 20$ meV) then suppressed to zero in the doped compounds ~\cite{jzhao2009,lwharriger2011,msliu2012,lwharriger2009,jtpark2012,hqluo2012b}. This is consistent with our previous INS measurements on the same sample at low energies and its more 2D-like nature in comparison with 122 systems ~\cite{txie2018a,sjiang2016b,cyjiang2020,sonora2017,fllovo2021}. For energies below 100 meV [$E=11\pm 1$, $21\pm 2$, $40\pm 4$, $60\pm 4$, $80\pm 10$ meV, Figs. 2 (a)-(e)], the spin excitations form transversely elongated ellipses centered around the in-plane AF ordering wave vectors $\mathbf{Q}_{\mathrm{AF}} = (\pm 1, 0)$ and (0, $\pm$1), and the peak intensity decreases with increasing energy. Such features are typically observed in the low energy spin excitations of those electron doped systems such as BaFe$_{2-x}$(Ni, Co)$_x$As$_2$, which are attributed to the anisotropic contributions from itinerant electrons due to the mismatched sizes between hole and electron pockets ~\cite{pdai2012,hfli2010,gstucker2012,msliu2012,hqluo2013}. When the energy increases to $E=100\pm 10$ meV,  $125\pm 10$ and  $140\pm 10$ meV [Figs. 2 (f), (g) and (h)], spin excitations start to split along the $K$-direction and disperse to the zone boundary $Q=(\pm1, \pm1)$. Finally, the spin excitations become blob-like at the AF zone boundary ($\pm$1, $\pm$1) when $E=210\pm 20$ meV [Fig. 2 (i)]. As the high-energy spin excitations are extremely weak, it looks two-fold due to limited statistics, which suppose to be four-fold like in $[H, K]$ plane in the pseudo-orthorhombic magnetic unit cell. To improve the statistics, we analysis the dispersion and intensity based on the four-fold data sets at low energies and two-fold data sets at high energies.

To firstly determine the rough dispersions of spin excitations in CaLa-112, we plot the background subtracted scattering for $E_i=450$, 250, 99 and 42 meV projected in the wave vectors $\mathbf{Q}=[1,K]$ and $[H, 0]$ and energy space. As clearly shown in Fig. 3 (a) and (c), the energy top for $K$ dispersion is about 200 meV, and for $H$ direction is about 150 meV, respectively. Such behaviors can be attributed to the anisotropic nearest-neighbor (NN) magnetic exchange coupling, namely $J_{1a}\neq J_{1b}$ in the effective Heisenberg model ~\cite{jzhao2009,lwharriger2011}. It is reasonable for such system with a monoclinic structure ($\alpha=90^{\circ}$, $\beta=91.4^{\circ}$)~\cite{sjiang2016a,sjiang2016b}. It can be also found that the low-energy spin excitations ($E<50$ meV) are very strong [Fig. 3 (b) and (d)]. For comparison, we also plot the spin wave dispersions of BaFe$_{2}$As$_{2}$ and CaFe$_{2}$As$_{2}$ as dashed lines in Fig. 3, which match overall with the excitations of CaLa-112.

\begin{figure*}[tbp]
\center
\includegraphics[width=1\textwidth]{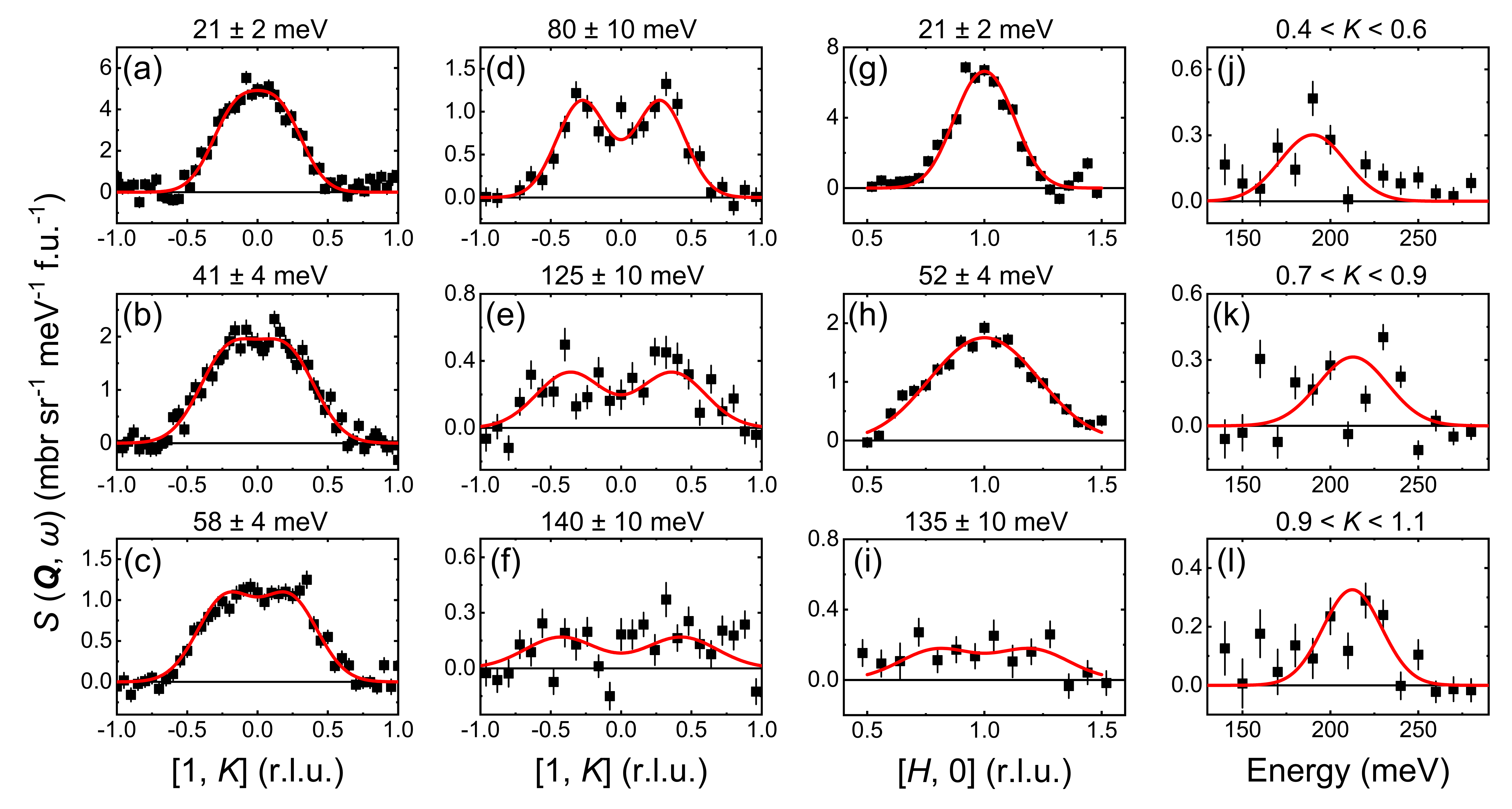}
\caption{(a-i) Constant-energy cuts in the spin excitations of CaLa-112 along the $[1,K]$ and $[H, 0]$ directions at different energies, where the wave vector integration ranges are $0.8<H<1.2$ for the $K$ cuts and $-0.2<K<0.2$ for the $H$ cuts.
 (j-l) Constant-$Q$ cuts in the spin excitations of CaLa-112 at wave vectors $Q= (1, 0.5)$, (1, 0.8), and (1, 1) with thickness $K\pm0.1$ and $0.8<H<1.2$. The red solid lines are Gaussian fitting results to determine the dispersions.}
\label{fig4}
\end{figure*}

To quantitatively illustrate the details of the in-plane dispersion of spin excitations in CaLa-112, we further show typical constant-energy cuts in Fig. 4 (a)-(i) at different energies along the $\mathbf{Q}=[1,K]$ and $[H, 0]$, respectively. The red solid lines are fitting results with two symmetric Gaussian functions. The spin excitations become initially incommensurate along the $[1, K]$ direction for $E=21 \pm 2$, $41 \pm 4$ and $58 \pm 4$ meV [Fig. 4 (a)-(c)], and clearly show a two-peak feature above 80 meV [Fig. 4 (d)-(f)]. The spin excitations seem to be commensurate along the $[H, 0]$ direction at low energies $E=21 \pm 2$ and $58 \pm 4$ meV [Fig. 4 (g)-(h)], and probably incommensurate at high energy $E=135 \pm 10$ meV with very weak intensity [Fig. 4 (i)]. Constant-$Q$ cuts at $Q= (1, 0.5)$, (1, 0.8), and (1, 1) are shown in Fig. 4 (j)-(l), where the peak positions correspond to the energy of $K$ dispersions approaching the AF zone boundary (open circles in Fig. 1(d)). The in-plane dispersions both along $[1,K]$ and $[H, 0]$ are determined from the fitting results of above 1D cuts, as shown in Fig. 1 (d) in comparison with the spin waves of BaFe$_{2}$As$_{2}$ and CaFe$_{2}$As$_{2}$.

Although it has been argued that the spin waves in BaFe$_{2}$As$_{2}$ are more appropriate to be described by an itinerant model when taking into account moderate electronic correlation effects~\cite{xylu2018}, a local-moment Heisenberg Hamiltonian with effective exchange couplings $J_{1a}$, $J_{1b}$ (nearest-neighbor, NN), $J_{2}$ (NNN) and $J_{c}$ (interlayer) can be also used to fit the spin waves in twinned samples by considering anisotropic couplings and dampings~\cite{jzhao2009,lwharriger2011}. Hence the dispersions are given by: $E(q)=\sqrt{A_q^2-B_q^2}$, where $A_q=2S[J_{1b}(\cos(\pi K)-1)+J_{1a}+2J_2+J_c+J_s]$, $B_q=2S[J_{1a}\cos(\pi H)+2J_2\cos(\pi H)\cos(\pi K)+J_c\cos(\pi L)]$, $J_s$ is the single ion anisotropy constant, and $q$ is the reduced wave vector away from the AF zone center. For BaFe$_{2}$As$_{2}$, $SJ_{1a}=59.2$ meV, $SJ_{1b}=-9.2$ meV, $SJ_{2}=13.6$ meV, $SJ_{c}=1.8$ meV, $SJ_{s}=0.084$ meV~\cite{lwharriger2011}, and other fittings give $SJ_{c}=0.22$ meV, $SJ_{s}=0.14$ meV~\cite{lwharriger2009,jtpark2012,hqluo2012b}. For CaFe$_{2}$As$_{2}$, $SJ_{1a}=49$ meV, $SJ_{1b}=-5.7$ meV, $SJ_{2}=18.9$ meV, $SJ_{c}=5.3$ meV, $SJ_{s}=0.063$ meV~\cite{jzhao2009}. As shown by the dashed lines Fig. 1(d), the dispersions of spin waves in BaFe$_{2}$As$_{2}$ and CaFe$_{2}$As$_{2}$ are similar under these two sets of exchange couplings. Except for some slight differences for $K$-dispersion below 50 meV, which were also observed in BaFe$_{2-x}$Ni$_x$As$_2$ ~\cite{hqluo2013}, the overall dispersion of CaLa-112 overlap with the spin waves of BaFe$_{2}$As$_{2}$ and CaFe$_{2}$As$_{2}$ within error bar. The in-plane dynamical spin-spin correlation length can be obtained from the Fourier transforms of the Gaussian fitting results of the constant-energy cuts in reciprocal space ($\xi=8\ln2/$FWHM, where FWHM is the full width at half maximum in the unit of \AA$^{-1}$) ~\cite{czhang2013}, which is about 10 \AA\ and similar to BaFe$_{2-x}$Ni$_x$As$_2$, too ~\cite{hqluo2013}.

Using the method described in the experimental details, we have calculated the energy dependence of dynamic local susceptibility $\chi^{\prime\prime}(E)$ by integrating the intensity of spin excitations in the AF Brillouin zone.  Due to the coupling between $E$ and $L$, we have to carefully choose the energy windows corresponding to $L\pm 0.5$ where $L=0.5, 1, 1.5, 2,...$, where the first Brillouin zone in $[H, K]$ plane is marked as the dashed boxes in Fig.2 and the thickness along $L$ direction should be $L\pm 0.5$ in such single-layered system. However, due to the $L$-independent signals in CaLa-112 ~\cite{txie2018a}, the $L$ integration range $L\pm 0.5$ should be equal to $L\pm 1$ by keeping in mind that the integration in DAVE and Horace software actually calculates the average signal over the integrated area. To avoid the interference from the unclean background at low energies ($E_i=$ 42, 73 meV), we perform such integration within a small area such as $0.7<H<1.3$ and $-0.3<K<0.3$, then normalize the data to the entire zone marked as the dashed boxes in Fig. 2. For high energy data ($E_i=$ 99, 250, 450 meV), we simply integrate the intensity in a large area after considering the twinning effect, which is identical to the dashed boxes in Fig. 2 as illustrated in the case of BaFe$_{2}$As$_{2}$ ~\cite{pdai2015}. Similar method was used in the calculation of $\chi^{\prime\prime}(E)$ in 122 system~\cite{mwang2013,msliu2012,hqluo2012b}. As shown in Fig. 1(b), the low energy spin susceptibility below 100 meV in CaLa-112 is indeed stronger than that in BaFe$_{2}$As$_{2}$ system, and the bandwidth, defined as the peak in energy dependence of $\chi^{\prime\prime}(E)$, is similar for these two compounds ($\sim$ 180 meV). The total fluctuating moments $\langle m^2 \rangle\approx 4.6\pm0.2 \mu_B^2$/Fe are slightly stronger than BaFe$_{2}$As$_{2}$, which are about $3.6 \mu_B^2$/Fe. These values are also slightly larger than the results of cuprates, in which $\langle m^2 \rangle> 1.9 \mu_B^2$/Cu ~\cite{jttranquada2014,mfujita2012,nsheadings2010,smhaydena1996,rcoldea2001}. Using the formula for magnetic moment of a spin $\langle m^2 \rangle=(g\mu_B)^2S(S+1)$ (where $g=2$), we can estimate that the CaLa-112 system has an effective spin $S\approx 0.68$, which likely corresponds to an $S=1/2$ magnetic ground state. These results are certainly different from the fully localized case, where $\langle m^2 \rangle=24 \mu_B^2$/Fe and $S=2$ under the $3d^6$ electronic configuration ~\cite{msliu2012}. Instead, it may be close to an $S=1$ ground state in the presence of itinerant electrons~\cite{qwang2016b}. Theoretically, the spin excitations in FeSCs may be alternatively described by a multi-orbital Hubbard-Hund model based on the pure itinerant picture as mentioned above~\cite{xylu2018,gong2022}, in which the intra- and inter-orbital on-site repulsion $U$ and the Hund's coupling $J_H$ are the effective parameters measuring the electronic correlation strength. Based on this picture, the density functional theory (DFT) combined with dynamical mean field theory (DMFT) calculation on the dynamic local susceptibility can predict the bandwidth of spin excitations in iron pnictides, as shown by the dashed dot line in Fig. 1(b) ~\cite{lwharriger2011,msliu2012,hqluo2013,dhu2016}. With increasing the interaction of electrons $U$, the Goldstone mode at low energy gains additional spectral weight. This could be a possible origin of the strong peak feature of $\chi^{\prime\prime}(E)$ around 20 meV, since the electronic correlation in CaLa-112 system may be enhanced by the involvement of As $4p$ orbitals in hybridization with the Fe $3d$ orbitals ~\cite{txie2018a,xliu2013,myli2015,zliu2016}. The itinerant nature of the low-energy spin excitations can be further confirmed by our previous polarized INS measurements, which suggest isotropic spin excitations in spin space~\cite{txie2018a}. From the comparison of spin excitations between CaLa-112 and BaFe$_{2-x}$Ni$_x$As$_2$ systems, we find another fact that the itinerancy of magnetism only affects the spin excitations below 100 meV in the electron doped iron pinictides, no matter they are suppressed or enhanced by dopings~\cite{mwang2013,msliu2012,hqluo2013}. Interestingly, in the electron doped iron chalcogenide FeSe$_{1-x}$Te$_x$, substitutions on Fe sites by Co, Ni and Cu strongly suppress the itinerancy, but the localization effects from Cu impurities enhance the low-energy spin excitations below 100 meV~\cite{jhwang2022}. Therefore, the energy scale $\sim$100 meV probably is a threshold to separate the dual contributions from local moments and itinerant electrons in FeSCs.

\section{Summary}
In summary, time-of-flight INS measurements are carried out to map the spin fluctuation spectrum in the 112-type FeSC Ca$_{0.82}$La$_{0.18}$Fe$_{0.96}$Ni$_{0.04}$As$_{2}$. The obtained results are compared with the parent compounds of 122-type FeSCs, while the in-plane dispersions, energy bandwidth and spin-spin correlation lengths are quite similar for both systems, the total fluctuating moments are stronger than BaFe$_{2}$As$_{2}$ probably due to more contributions from itinerant electrons at low energies. Therefore, the magnetic exchange couplings should be similar between 112 and 122 systems but the fluctuating effective spin is not, even though they have different magnetic patterns and lattice symmetries.

\section*{Acknowledgements}

This work is supported by the National Key Research and Development Program of China (Grants No. 2018YFA0704200, No. 2017YFA0303100, and No. 2017YFA0302900), the National Natural Science Foundation of China (Grants No. 11822411, No. 11961160699, and No. 12061130200), the Strategic Priority Research Program (B) of the CAS (Grants No. XDB25000000 and No. XDB33000000) and K. C. Wong Education Foundation (GJTD-2020-01). H. L. is grateful for the support from the Youth Innovation Promotion Association of CAS (Grant No. Y202001) and Beijing Natural Science Foundation (Grants No. JQ19002). Work at Oak Ridge National Laboratory (ORNL) was supported by the U.S. Department of Energy (DOE), Office of Science, Basic Energy Sciences, Materials Science and Engineering Division. This work is based on inelastic neutron scattering experiments performed at the Materials and Life Science Division of J-PARC, Ibaraki, Japan (Proposal No. 2017B0058).

\end{document}